\documentclass[final]{siamltex}

\usepackage{graphicx}
\usepackage{color}
\usepackage{amssymb,amsmath,amsfonts}
\usepackage{bm}
\usepackage{multicol}

\RequirePackage[pdftex,pdfpagemode=none, pdftoolbar=true,
  pdffitwindow=true,pdfcenterwindow=true]{hyperref}

\definecolor{darkgreen}{rgb}{0,0.5,0}

\usepackage{float}

\makeatletter
\let\IfDraftVersion\if@draftclsmode
\makeatother

\def\Gb{{\protect\mathbf{G}}}

\def\Nb{{\protect\mathbf{N}}}

\def\Tb{{\protect\mathbf{T}}}

\def\xb{{\protect\mathbf{x}}}
\def\yb{{\protect\mathbf{y}}}

\def\Rbmit{{\protect\bm{R}}}
\def\Tbmit{{\protect\bm{T}}}

\def\Vbmit{{\protect\bm{V}}}

\def\Xbmit{{\protect\bm{X}}}
\def\Ybmit{{\protect\bm{Y}}}
\def\Zbmit{{\protect\bm{Z}}}

\def\zerob{{\protect\bm{0}}}

\DeclareMathOperator{\atantwoactual}{atan2}

\def\atantwo(#1,#2){{\tan^{-1}(#1/#2)}}

\def\T{{\scriptscriptstyle\rm T}}   
\let\humlaut=\H
\def\H{\ifmmode{\scriptscriptstyle\rm H}\else\humlaut\fi}   

\def\by{\ifmmode $\hbox{-by-}$\else \leavevmode\hbox{-by-}\fi}
\def\sqrtm1{{\sqrt{\!-1}}}
\let\(=\langle
\let\)=\rangle

\def\texthalf{{\textstyle{1\over2}}}

\def\Cerveny{\v{C}erven\'y}

\makeatletter

\newlength\eqncolsep
\let\eqnarray@LaTeX\relax
\let\eqnarray@fleqn@fixed\relax
\def\mathindent{\@centering}%
\def\t@{to}%
\def\halignt@{\halign\t@}%

\long\def\eqnarray@fleqn@hfil{%
 \stepcounter{equation}\def\@currentlabel{\p@equation\theequation}%
 \global\@eqnswtrue\m@th\global\@eqcnt\z@
 \tabskip\mathindent
 \let\\=\@eqncr
 \setlength\abovedisplayskip{\topsep}%
 \ifvmode\addtolength\abovedisplayskip{\partopsep}\fi
 \addtolength\abovedisplayskip{\parskip}%
 \setlength\belowdisplayskip{\abovedisplayskip}%
 \setlength\belowdisplayshortskip{\abovedisplayskip}%
 \setlength\abovedisplayshortskip{\abovedisplayskip}%
 $$%
 \everycr{}%
 \halignt@\linewidth\bgroup\hfil
  \hskip\@centering$\displaystyle\tabskip\z@skip{##}$\@eqnsel
  &\global\@eqcnt\@ne
   \hskip\tw@\eqncolsep
   \hfil${{}##{}}$\hfil
  &\global\@eqcnt\tw@
   \hskip\tw@\eqncolsep
   $\displaystyle{##}$\hfil\tabskip\@centering
  &\global\@eqcnt\thr@@\hb@xt@\z@\bgroup\hss##\egroup
   \tabskip\z@skip
  \cr
}%

\long\def\dbleqnarray{%
 \stepcounter{equation}\def\@currentlabel{\p@equation\theequation}%
 \global\@eqnswtrue\m@th\global\@eqcnt\z@
 \tabskip\mathindent
 \let\\=\@eqncr
 \setlength\abovedisplayskip{\topsep}%
 \ifvmode\addtolength\abovedisplayskip{\partopsep}\fi
 \addtolength\abovedisplayskip{\parskip}%
 \setlength\belowdisplayskip{\abovedisplayskip}%
 \setlength\belowdisplayshortskip{\abovedisplayskip}%
 \setlength\abovedisplayshortskip{\abovedisplayskip}%
 $$%
 \everycr{}%
 \halignt@\linewidth\bgroup\hfil
  \hskip\@centering$\displaystyle\tabskip\z@skip{##}$\@eqnsel
  &\global\@eqcnt\@ne
   \hskip\tw@\eqncolsep
   \hfil${{}##{}}$\hfil
  &\global\@eqcnt\tw@
   \hskip1\eqncolsep
   $\displaystyle{##}$\hfil\tabskip\@centering
  &\qquad\hskip\@centering\hfil$\displaystyle\tabskip\z@skip{##}$\@eqnsel
  &\global\@eqcnt\@ne
   \hskip\tw@\eqncolsep
   \hfil${{}##{}}$\hfil
  &\global\@eqcnt\tw@
   \hskip\tw@\eqncolsep
   $\displaystyle{##}$\hfil\tabskip\@centering
  &\global\@eqcnt\thr@@\hb@xt@\z@\bgroup\hss##\egroup
   \tabskip\z@skip
  \cr
}%

\long\def\debpeqnarray{%
 \stepcounter{equation}\def\@currentlabel{\p@equation\theequation}%
 \global\@eqnswtrue\m@th\global\@eqcnt\z@
 \tabskip\mathindent
 \let\\=\@eqncr
 \setlength\abovedisplayskip{\topsep}%
 \ifvmode\addtolength\abovedisplayskip{\partopsep}\fi
 \addtolength\abovedisplayskip{\parskip}%
 \setlength\belowdisplayskip{\abovedisplayskip}%
 \setlength\belowdisplayshortskip{\abovedisplayskip}%
 \setlength\abovedisplayshortskip{\abovedisplayskip}%
 $$%
 \everycr{}%
 \halignt@\linewidth\bgroup\hfil
  \hskip\@centering$\displaystyle\tabskip\z@skip{##}$\@eqnsel
  &\global\@eqcnt\@ne
   \hskip\tw@\eqncolsep
   \hfil${{}##{}}$\hfil
  &\global\@eqcnt\tw@
   \hskip\tw@\eqncolsep
   $\displaystyle{##}$\hfil\tabskip\z@skip
  &\quad\hskip\@centering\hfil$\displaystyle\tabskip\z@skip{##}$\@eqnsel
  &\global\@eqcnt\@ne
   \hskip\tw@\eqncolsep
   \hfil${{}##{}}$\hfil
  &\global\@eqcnt\tw@
   \hskip\tw@\eqncolsep
   $\displaystyle{##}$\hfil\tabskip\@centering
  &\global\@eqcnt\thr@@\hb@xt@\z@\bgroup\hss##\egroup
   \tabskip\z@skip
  \cr
}%

\makeatother

\makeatletter

\newif\ifmathtomb \mathtombfalse

\def\tombstone{\unskip\penalty50   
  \hskip 0pt plus-1fill \null\nobreak\hskip 0pt plus1fill
  \enskip \vrule width.3333em height.7em depth.2em
  \ifmmode \global\mathtombtrue \else \global\mathtombfalse \fi}

  {\futurelet\next\hpr@oftext}
  {\ifmathtomb \else \tombstone \fi \widowpenalty=10000  
   \par \ifmathtomb \else \addvspace{\medskipamount}\fi \global\mathtombfalse}
\def\hpr@oftext{\ifx\next[\let\temp\ohpr@@ftext\else\let\temp\hpr@@ftext\fi\temp}
\def\hpr@@ftext{\beginhpr@@f{Proof}}
\def\ohpr@@ftext[#1]{\beginhpr@@f{#1}}
\def\beginhpr@@f#1{\par \addvspace{\bigskipamount}%
  \noindent{\bf #1:\enspace}\ignorespaces }

\def\intrgeomspr{\mathayn}    
\def\dotintrgeomspr{\mathdotayn}    

\makeatletter
\let\endabstract@original\endabstract
\makeatother

\usepackage{arabtex}

\makeatletter
\let\endabstract\endabstract@original
\makeatother

\novocalize

\DeclareSymbolFont{arabmath}{U}{xnsh}{m}{s}
\DeclareSymbolFont{arabmathbold}{U}{xnsh}{bx}{s}

\DeclareMathSymbol{\alif}{0}{arabmath}{"40}
\DeclareMathSymbol{\ayn}{0}{arabmath}{"A8}
\DeclareMathSymbol{\alifb}{0}{arabmathbold}{"40}
\DeclareMathSymbol{\aynb}{0}{arabmathbold}{"A8}

\def\mathayn{{\kern.1em\raise.3ex\hbox{$\ayn$}\kern.05em}}
\def\mathaynb{{\kern.1em\raise.3ex\hbox{$\aynb$}\kern.05em}}
\def\mathdotayn{{\kern.1em\raise.3ex\hbox{$\dot\ayn$}\kern.05em}}
\def\mathdotaynb{{\kern.1em\raise.3ex\hbox{$\dot\aynb$}\kern.05em}}

\title{On Gaussian beams described by Jacobi's equation}

\begin{document}

\author{Steven T. Smith}\thanks{MIT Lincoln Laboratory; 244 Wood
  Street; Lexington MA 02420 USA;
  \href{mailto:stsmith@ll.mit.edu}{stsmith@\penalty50ll.mit.edu}. This
  work is sponsored by the Department of the United States Navy under
  Air~Force contract FA8721-05-C-0002.  Opinions, interpretations,
  conclusions, and recommendations are those of the author and are not
  necessarily endorsed by the United States Government.}

\date{22 March 2013}

\maketitle

\begin{abstract} Gaussian beams describe the amplitude and phase of
rays and are widely used to model acoustic propagation. This paper
describes four new results in the theory of Gaussian beams. (1)~A new
version of the \Cerveny\ equations for the amplitude and phase of
Gaussian beams is developed by applying the equivalence of
Hamilton-Jacobi theory with Jacobi's equation that connects Riemannian
curvature to geodesic flow.  Thus the paper makes a fundamental
connection between Gaussian beams and an acoustic channel's so-called
intrinsic Gaussian curvature from differential geometry. (2)~A new
formula $\pi(c/c'')^{1/2}$ for the distance between convergence zones
is derived and applied to the Munk and other well-known
profiles. (3)~A class of ``model spaces'' are introduced that connect
the acoustics of ducting\slash divergence zones with the channel's
Gaussian curvature ${K=cc''-(c')^2}$. The model SSPs yield constant
Gaussian curvature in which the geometry of ducts corresponds to great
circles on a sphere and convergence zones correspond to antipodes. The
distance between caustics $\pi(c/c'')^{1/2}$ is equated with an ideal
hyperbolic cosine SSP duct. (4)~An intrinsic version of \Cerveny's
formulae for the amplitude and phase of Gaussian beams is derived that
does not depend on an extrinsic, arbitrary choice of coordinates such
as range and depth. Direct comparisons are made between the
computational frameworks used by the three different approaches to
Gaussian beams: Snell's law, the extrinsic Frenet-Serret formulae, and
the intrinsic Jacobi methods presented here. The relationship of
Gaussian beams to Riemannian curvature is explained with an overview
of the modern covariant geometric methods that provide a general
framework for application to other special cases.\end{abstract}



\pagestyle{myheadings}
\thispagestyle{plain}
\markboth{Steven T. Smith}{On Gaussian beams described by Jacobi's equation}

\section*{Notation and SI Units} \ 

\setlength{\columnsep}{15pt}

\begin{multicols}{2}
{\scriptsize
\tabskip=0pt plus1fil
\def\siskip{\hskip 0.5em plus.5em minus.25em}
\halign to\hsize{$#$\hfil\tabskip=0.75em minus.5em&#\hfil\tabskip=0pt plus1fil\hidewidth&\siskip\hfil#\tabskip=0pt\cr
t&Time&[s]\cr
s&Arclength&[m]\cr
T&Ray travel-time&[s]\cr
\xb=(r,z)^\T&Range and depth&[m]\cr
c(z)&Sound speed profile&[m/s]\cr
c'={d\over dz}c(z)&Derivative of~SSP&[1/s]\cr
\dot\xb={d\over dt}\xb&Velocity vector&[m/s]\cr
\Tb={d\over ds}\xb&Unit tangent vector&[1]\cr
\Nb&Unit normal vector&[1]\cr
g(\dot\xb,\dot\xb),\ g_{ij}&Riemannian metric&[1, $\rm s^2\!/m^2$]\cr
\theta_0&Initial elevation angle&[rad]\cr
L&Lagrangian function&[1]\cr
H&Hamiltonian function&[1]\cr
\varkappa&Extrinsic curvature&[1/m]\cr
\delta q&Ray distance along~$\Nb$&[m]\cr
\delta p=\partial T/\partial\delta q&Conjugate momentum of~$\delta q$&[s/m]\cr
\noalign{\penalty-500}
\delta\tilde q=c^{-1}\,\delta q&Travel-Time along~$\Nb$&[s]\cr
\delta\tilde p=c\,\delta p&Conjugate momentum of~$\delta\tilde q$&[1]\cr
\noalign{\penalty-500}
q=\lim{\delta q\over\delta\theta_0}&Extrinsic geom.\ spreading&[m/rad]\cr
p=\lim{\delta p\over\delta\theta_0}&Conj.\ momentum of~$q$&[s/m/rad]\cr
\noalign{\penalty-500}
\tilde q=c^{-1}q&Intrinsic geom.\ spreading&[s/rad]\cr
\tilde p=cp&Conjugate momentum of~$\tilde q$&[1/rad]\cr
\noalign{\penalty-500}
\intrgeomspr=\tilde q&Intrinsic geom.\ spreading&[s/rad]\cr
\dotintrgeomspr=(d/dt)\intrgeomspr&First derivative of~$\intrgeomspr$&[1/rad]\cr
\noalign{\penalty-500}
\delta t_{\rm e}=\texthalf{p\over q}\,\delta q^2&Extrinsic ray tube phase&[s]\cr
\delta t_{\rm i}=\texthalf{\dotintrgeomspr\over\intrgeomspr}\,\delta\tilde q^2&Intrinsic ray tube phase&[s]\cr
\Tbmit=\dot\xb&Ray tangent vector&[m/s]\cr
\Vbmit=\intrgeomspr\Ybmit&Variation vector&[m/rad]\cr
\Ybmit&Unit parallel vector&[m/s]\cr
K&Gaussian curvature&[$1/{\rm s}^2$]\cr
\Rbmit(\Vbmit,\Tbmit)&Riemannian curvature&[1]\cr
R^i{}_{jkl}&R.\ curvature coefficients&[$1/{\rm m}^2$]\cr
\nabla_\Tbmit&\omit Covariant differentiation&[1]\cr
\Gamma^k_{ij}&Christoffel symbols&[1/m]\cr}}
\end{multicols}

\section{\label{sec:intro}Introduction}

This paper uses Jacobi's equation to derive new formulae for the
geometric spreading loss and phase through Gaussian beams, and thus
provides an alternate method for paraxial ray
tracing.~\cite{Bos09,Bos10,Cerveny87,Cerveny01,Jobert87,Keller83,Nair91,Porter87,Jensen11}
The new formulation, though mathematically equivalent to well-known
expressions, provides new geometric insight into the physical and
intrinsic geometric characteristics of Gaussian beams and their
relationship to the Gaussian and Riemannian curvature of the
propagation medium. Thus the expressions introduced in this paper
establish the connection between Gaussian beams and the intrinsic
geometry of the propagation medium. Four new geometrically-motivated
ideas for ducting are presented: (1)~the \Cerveny\ equations for the
amplitude and phase of Gaussian beams are expressed in a new form
using the equivalence of the Hamilton-Jacobi equations that involves
the Hamiltonian and Jacobi's equation that involves Riemannian
curvature; (2)~a new formula $\pi(c/c'')^{1/2}$ for the distance
between convergence zones is derived, where $c(z)$ is the sound speed
profile (SSP); (3)~the intrinsic geometry of acoustic ducting is shown
to be equivalent to great circles on a sphere with convergence zones
corresponding to antipodes; (4)~a coordinate-free ``intrinsic''
version of \Cerveny's formulae for the amplitude and phase of Gaussian
beams is presented.

Paraxial ray methods are generally known as ``Gaussian beams'' because
each ray is treated as representing a volume or ray tube in which the
ray's amplitude and phase in the transverse plane perpendicular to the
ray's tangent is determined by a Gaussian density. Transversely along
the ray, Jacobi's equation determines the geometric spreading loss,
expressed using Riemannian or sectional curvatures, or, in the case of
$2$\hbox{-}d rays, the Gaussian curvature. Tangentially, the relative
time lag of nearby rays determines the phase of the Gaussian beam,
which themselves are described by complex solutions $Ae^{j\,\varphi}$
to the Hamilton-Jacobi equations, where $A$ is the beam amplitude,
$\varphi$ is the beam phase, and ${j=\sqrt{-1}}$.  Therefore, the name
``Gaussian beam'' is highly suitable because Gaussian beams are
completely determined by Gauss's eponymous curvature.

One interesting example of a new physical insight derived from
Jacobi's equation is a simple formula for the distance between
caustics: it is shown that the half-wavelength distance is about
$\pi(c/c'')^{1/2}$, a quantity that depends entirely on the SSP. It
will be proved that the distance between convergence zones for a Munk
profile with parameter $\epsilon$ and scaled depth~$W$~meters is
about~$\pi\epsilon^{-1/2}W$~meters. For SSPs with an idealized
hyperbolic cosine profile ${c(z)=c_0\cosh(z-z_0)/W}$, this distance is
shown to equal exactly~$\pi W$~meters for~all rays.  Caustics arise
with positive Gaussian curvature; when the curvature is negative, rays
diverge and Jacobi's equation quantifies their divergence, or
transmission loss. For~example, for linear SSPs with slope~$c'$ the
geometric spreading at time~${t\lessapprox c'^{-1}}$ is about
${ct+{1\over 6}c(c')^2t^3}$.  These results are both a direct
consequence of the fact that the Gaussian curvature of the propagation
medium equals ${K=cc''-(c')^2}$. Classifying the SSP by its Gaussian
curvature will allow for the introduction of model spaces for
convergent ducts whose curvature is constant positive, divergence
zones with constant negative curvature, and simple non-refractive
spreading with vanishing curvature.  Another feature arising from this
work is an accounting of the additional spreading loss for either
reflected or transmitted rays at an interface, at which point the
Gaussian curvature is not defined.

Direct comparisons are made between the computational frameworks
derived from the three different approaches to Gaussian beams:
(1)~Snell's
law,~\cite{Brekhovskikh80,Pierce89,Brekhovskikh99,Brekhovskikh03,Clay77,Jensen11,Westwood87}
(2)~a variant of the extrinsic Frenet-Serret established by
\v{C}erven\'y and colleagues,~\cite{Cerveny87,Cerveny01} and (3)~the
new intrinsic methods presented here.
Bergman,~\cite{Bergman05a,Bergman05b,Bergman06} apparently the first
to recognize the application of Jacobi's equation to ray tracing,
recently adopted methods from General Relativity to address the
problem of computing ray amplitudes in a relativistic acoustic
field. The non-relativistic intrinsic results developed in this paper
are equivalent to many of Bergman's if one uses the space-like part of
his pseudo-Riemannian Lorentz metric.

It is perhaps noteworthy that Jacobi's equation and its full
implications for Gaussian beams has, apparently, not yet appeared in
the acoustical literature.  This lacuna might be defended by a few
historical observations. Lord Rayleigh was initially dismissive of the
practical applications for acoustic refraction, noting in 1877
``almost the only instance of acoustical refraction, which has
practical interest, is the deviation of sonorous rays from rectilinear
course due to the heterogeneity of the
atmosphere.''~\cite{Rayleigh45} Though the foundations of
non-Euclidean geometry had been established by this time, differential
and Riemannian geometry remains relatively little known in engineering
applications to this day, in spite of the fact that covariant analysis
is the natural approach to many physical and engineering problems, as
will be seen in its application to Gaussian beams here.

In section~\ref{sec:rayamp2} the ray equations developed using the
classical Euler-Lagrange, Frenet-Serret, and Hamilton-Jacobi
formulations, followed by the amplitude and phase along a specific ray
using \Cerveny's Hamilton-Jacobi approach. This standard development
is then recast using an intrinsic parameterization that will be shown
equivalent to Jacobi's equation. Section~\ref{sec:geometric-spreading}
develops an intrinsic formulation of Gaussian beams based on Jacobi's
equation and explores the physical consequences of this approach,
including a computation of the distance between convergence zones and
a classification of SSPs using ``model spaces'' of constant Gaussian
curvature.


\section{\label{sec:rayamp2}Gaussian Beams in Horizontally Stratified Isotropic Media}

The simplest and most frequently encountered case of acoustic rays in
a horizontally stratified isotropic medium will be analyzed.  As
usual, denote the three spatial coordinates by the variables $x$, $y$,
and~$z$, time by~$t$, the spatial infinitesimal arclength by
$ds^2=dx^2+dy^2+dz^2$, and the depth-dependent SSP by~$c(z)$,
differentiable except at a discrete set of points where either $c(z)$
itself is discontinuous (e.g.\ Snell's law), or its first derivative
$c'(z)=dc/dz$ is discontinuous (e.g.\ method of images at a
boundary). Paths $\xb(t) =\bigl(r(t),\phi(t),z(t)\bigr)^\T$ are
expressed using cylindrical coordinates, with $r=\sqrt{\mathstrut
  x^2+y^2}$, $\phi =\atantwo(y,x)$, and $z$ (pointing down), so that
$ds^2=dr^2+r^2\,d\phi^2+dz^2$.  The physical solution is independent
of these coordinates.  The time required to travel along an arbitrary
continuous path equals \begin{equation} T[\xb(t)] =\int dt =\int
  {1\over c(z)}{ds\over dt}dt =\int c^{-1}(z)(\dot r^2+r^2\dot\phi^2
  +\dot z^2)^{1/2}\,dt.\label{eq:pathtime}\end{equation} where
$\dot\xb(t)=d\xb/dt$ and~${ds/dt=c(z)}$. The Fermat metric is
represented by the quadratic function $g(\dot\xb,\dot\xb)
=c^{-2}(z)(\dot r^2 +r^2\dot\phi^2+\dot z^2)$, whose square-root
appears in Eq.~(\ref{eq:pathtime}). This Riemannian metric has
coefficients ${\Gb =(g_{ij})}$ and induces an inner product
$\langle\dot\xb,\dot\yb\rangle =g(\dot\xb,\dot\yb) =c^{-2}(z)(\dot
r_1\dot r_2 +r^2\dot\phi_1\dot\phi_2 +\dot z_1\dot z_2)$ and norm
$\|\dot\xb\|^2 =g(\dot\xb,\dot\xb)$ on the space of tangent vectors
$(\dot r,\dot\phi,\dot z)^\T$ at each point $(r,\phi,z)^\T$.

\begin{figure}[t]
\begin{multicols}{2}
\begin{figure}[H]
\centerline{\includegraphics[width=\columnwidth]{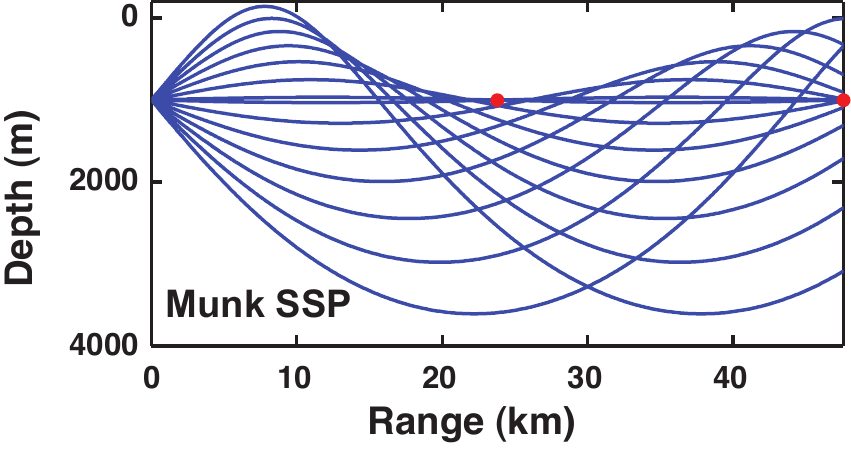}}
\caption{Rays for the Munk SSP~\cite{Jensen11} with with parameter
  ${\epsilon=0.00737}$, scaled depth ${\bar z=(z-z_0)/W}$ with
  ${W=650\,}$m, and (nonconstant) Gaussian curvature ${K=\epsilon
    c_0^2/W^2}$ at~$z_0$, yielding convergence zones at about every
  ${\pi\epsilon^{-1/2}W=23.8\,}$km (Theorem~\ref{thm:munkdist}) as
  illustrated by the {\color{red}red}
  dots~(\/{\color{red}$\bullet$}). The rays are computed using a
  standard 4th order Runge-Kutta ODE solver.\label{fig:munk-ssp}}
\end{figure}
\begin{figure}[H]
\centerline{\includegraphics[width=0.75\columnwidth]{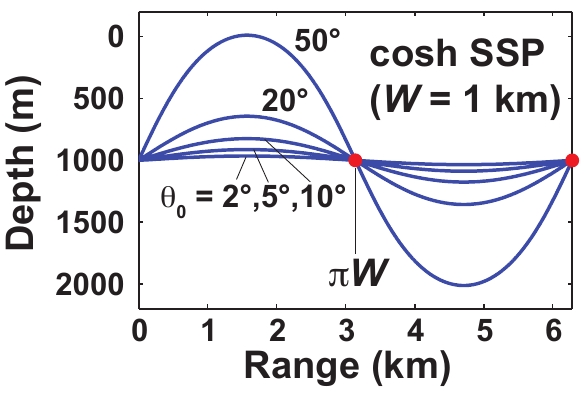}}
\caption{Rays for the hyperbolic cosine SSP $c(z)=c_0\cosh({z-z_0})/W$
  with $z_0=W=1\,$km. This SSP yields a space of constant positive
  curvature, resulting in a duct with convergent rays and caustics
  independent of any initial elevation angle~$\theta_0$ at ranges ${\pi
    W=3.14159}\ldots\,$km (Theorem~\ref{thm:cosh-ssp}), illustrated by
  the {\color{red}red} dots~(\/{\color{red}$\bullet$}). The rays are
  computed using a standard 4th order Runge-Kutta ODE
  solver.\label{fig:cosh-ssp}}
\end{figure}
\end{multicols}
\vspace{-12pt}
\end{figure}

Before the main results of the paper involving the geometry of
acoustic ducting and divergence are presented, a concise background of
ray theory is provided in Subsections
\ref{sec:ray-euler-lagrange-frenet-serret}
and~\ref{sec:ray-hamilton-jacobi} using the Euler-Lagrange,
Frenet-Serret, and Hamilton-Jacobi formulations.
Subsection~\ref{sec:paraxial-ray-hamilton-jacobi} derives the
amplitude and phase along a specific ray using \Cerveny's
Hamilton-Jacobi approach.  \Cerveny's equations are recast in
section~\ref{sec:paraxial-ray-intrinsic} using an intrinsic
parameterization that will be shown in
section~\ref{sec:geometric-spreading} to yield Jacobi's equation
expressed in form that yields geometric insight into acoustic ducting.

\subsection{\label{sec:ray-euler-lagrange-frenet-serret}Ray Equations: Euler-Lagrange and Frenet-Serret}

Fermat's principle implies that rays satisfy the Euler-Lagrange
equation, ${(d/dt)(\partial L/\partial\dot\xb) -(\partial
  L/\partial\xb)}=\zerob$, with either initial conditions or boundary
conditions for eigenrays. Attention is restricted to the $(r,z)$-plane
because radial symmetry implies that $\phi(t)\equiv\phi_0$. The
Euler-Lagrange equations with Lagrangian $L(t;r,z;\dot r,\dot z)
=c^{-1}(z)(\dot r^2+\dot z^2)^{1/2}$ and radial symmetry
${\dot\phi\equiv0}$ yields the well-known differential ray
(Christoffel) equations \begin{equation} \ddot r -2(c'/c)\dot r\dot
  z=0,\qquad \ddot z +(c'/c)(\dot r^2-\dot
  z^2)=0.\label{eq:rayrz2} \end{equation} Figs.\ \ref{fig:munk-ssp}
and~\ref{fig:cosh-ssp} illustrate computed rays determined by the Munk
and hyperbolic cosine SSPs. Parameterization by travel-time~$t$ is
said to be ``natural'' or ``intrinsic'' because rays minimize
travel-time. It is oftentimes computationally convenient to
parameterize rays by ``extrinsic'' arclength $ds=({dr^2+dz^2})^{1/2}$,
in which case Eq.~(\ref{eq:rayrz2}) becomes the first Frenet-Serret
formula $(d^2/ds^2)(r,z)^\T =(c'/c)(dr/ds)(dz/ds,-dr/ds)^\T$,
i.e.\ $d\Tb/ds =\varkappa \Nb$, where $\Tb=(dr/ds,dz/ds)^\T$ is the
ray's tangent vector, $\Nb=(dz/ds,-dr/ds)^\T$ is its normal (always
defined in the same direction), and $\varkappa = (c'/c)\*(dr/ds)
=-c_{\rm n}/c$ is the ray's extrinsic curvature, and $c_{\rm
  n}=(\partial c/\partial\xb){\cdot}\Nb =-c'(dr/ds)$ is the first
derivative of the SSP~$c$ in the normal direction.  The quadratic
coefficients for the tangent vector (first derivative) terms that
appear in the ray equations [Eq.~(\ref{eq:rayrz2})] are called
Christoffel symbols of the second
kind,~\cite{Cheeger75,Helgason78,Rund59,Spivak99} and are crucial in
quantifying the amplitude and phase along the ray caused by geometric
spreading. A continuous version of Snell's
law~\cite{Brekhovskikh80,Brekhovskikh03,Clay77,Jensen11,Westwood87}
yields the integral solution \begin{equation} r(z;\theta_0)
  =\int_{z_0}^z{ac(z_\prime)\over\bigl(1-a^2c^2(z_\prime)\bigr)^{1/2}}\,dz_\prime,
  \label{eq:snells-law} \end{equation} in which
${a=c^{-1}(z_0)\cos\theta_0}$ is the Snell invariant with initial
conditions $\bigl(r(0),z(0)\bigr)=(0,z_0)$ and $\bigl(\dot r(0), \dot
z(0)\bigr) =c(z_0)(\cos\theta_0,-\sin\theta_0)$.

\subsection{\label{sec:ray-hamilton-jacobi}Ray Equations: Hamilton-Jacobi Equation}

Rays satisfy the canonical form of the Euler-Lagrange equations
${z'=\partial H/\partial \zeta}$, ${\zeta'= -\partial H/\partial z}$
with Hamiltonian $H(r;z;\zeta)=\zeta z'-L(r;z;z')$ and~$\zeta
=\partial L/\partial z'$. The travel-time $T$ of
Eq.~(\ref{eq:pathtime}) satisfies the Hamilton-Jacobi equation
$(\partial T/\partial r) +H(r;z;\partial T/\partial z) =0$, which
reduces to the well-known eikonal equation $(\partial T/\partial r)^2
+(\partial T/\partial z)^2 =c^{-2}(z)$ for the Hamiltonian
$H(r;z;\zeta) =-\bigl(c^{-2}(z) -\zeta^2\bigr)^{1/2}$ corresponding to
the Lagrangian in section~\ref{sec:ray-euler-lagrange-frenet-serret}.

\subsection{\label{sec:paraxial-ray-hamilton-jacobi}Paraxial Ray Equations: Hamilton-Jacobi Form}

\begin{figure}[t]
\centerline{\includegraphics[width=0.5\columnwidth]{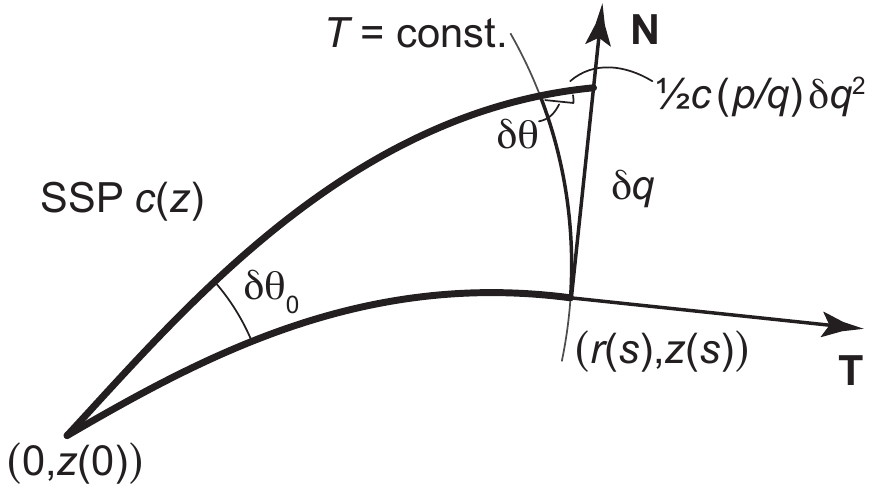}}
\caption{Ray tube in a horizontally stratified medium. The azimuthal
  component, not shown, is perpendicular to the page. All distances in
  the figure represent actual, extrinsic Euclidean distance. The
  extrinsic distance between nearby rays along the normal vector $\Nb$
  is denoted as~$\delta q$, and the nearby ray's additional length
  equals ${1\over2}c(\delta p/\delta q)\,\delta q^2
  ={1\over2}c(p/q)\,\delta q^2$, where $q=\lim\delta q/\delta\theta_0$
  is the extrinsic geometric spreading, and $\delta p=\partial
  T/\partial\delta q$ and $p=\partial T/\partial q$ are the conjugate
  momenta corresponding to~$\delta q$ and~$q$.\label{fig:raytube}}
\end{figure}

Ray amplitude is determined by the spreading of nearby rays, and phase
away from the ray is determined by time differences, so that rays are
viewed as tubes or beams possessing both amplitude and phase rather
than the skeletal objects determined by Eq.~(\ref{eq:rayrz2}). As
first established by \Cerveny\ and
colleagues,~\cite{Cerveny87,Cerveny01,Jobert87,Porter87,Jensen11} the
equations for the amplitude and phase along a ray tube are given by
the canonical equations after a convenient change of variables
involving the ray itself. Let $s$ be the arclength along the ray and
let $\delta q$ be the small or infinitesimal distance away from the
ray, measured perpendicularly along the normal~$\Nb$ at arclength~$s$
such that \begin{equation} \begin{pmatrix} r(s,\delta q)\\z(s,\delta
    q)\end{pmatrix} =\begin{pmatrix} r(s)\\z(s)\end{pmatrix} +\delta
  q\begin{pmatrix}
  dz/ds\\-dr/ds\end{pmatrix}, \label{eq:qdef}\end{equation} and let
\begin{equation}\delta p=\partial T/\partial \delta q\label{eq:deltap}\end{equation}
be the small or infinitesimal derivative of travel-time w.r.t.~$\delta
q$ from the ray along the line~$\Nb$, all illustrated in
Fig.~\ref{fig:raytube}. Thus, $\delta q$ quantifies the spread of
nearby rays, and therefore determines the ray's amplitude, and as
shown by \Cerveny~\cite{Cerveny01} the conjugate momentum $\delta p$
appears in the difference in time-of-travel nearby rays along~$\Nb$
and thus determines the ray's phase. Following
\Cerveny~\cite{Cerveny01} (also see Wolf and Kr\"otzsch~\cite{Wolf95})
Hamilton-Jacobi theory is used to determine the governing equations
for $\delta q$ and~$\delta p$.

By the chain rule $\partial T/\partial(s,\delta q) =\partial
T/\partial(r,z){\cdot}J$, where $J=\partial (r,z)/\partial(s,\delta q)
=\bigl(h\Tb,\Nb\bigr)$ is the Jacobian matrix along the ray at
distance~$s$ at ${\delta q=0}$ with scale factor $h(s,\delta q)
=1-\bigl((c'/c)(dr/ds)\bigr)|_{\delta q=0}\,\delta q =1+(c_{\rm
  n}/c)|_{\delta q=0}\,\delta q$. The eikonal equation expressed in
the ray-centered coordinates $s$ and~$\delta q$ is $h^{-2}(s,\delta
q)(\partial T/\partial s)^2 +(\partial T/\partial \delta q)^2
=c^{-2}(s,\delta q)$, corresponding to the Hamiltonian
  $H(s;\delta q;\delta p) =-h(s,\delta q)\*\bigl({c^{-2}(s,\delta q)
  -\delta p^2}\bigr)^{1/2}$. For small $\delta q$ and~$\delta p$ up to
  second-order, \begin{eqnarray} {1\over c(s,\delta q)} &=&{1\over
      c_s} -{c_{\rm n}\over c_s^2}\,\delta q -{1\over2}\left({c_{\rm
        nn}\over c_s^2} -2{c_{\rm n}^2\over c_s^3}\right)\,\delta q^2
    +\cdots,\label{eq:csqinv}\\ H(s;\delta q;\delta p) &=&
    -hc^{-1}(s,\delta q)(1-\texthalf c_s^2\,\delta p^2) +\cdots
    \\ \omit\hbox to0pt{$\displaystyle\quad{} = -c_s^{-1} +{c_{\rm
          nn}\over 2c_s^2}\,\delta q^2 +{1\over2}c_s\,\delta p^2
      +\cdots$,\hss}\hfill\label{eq:Hsqsmall} \end{eqnarray} where
  $c_s=c\bigl(z(s)\bigr)$ and $c_{\rm
    nn}=c''\bigl(z(s)\bigr)(dr/ds)^2$ is the second derivative of the
  SSP in the direction of the ray normal.

Applying Hamilton's equations of motion ${d\delta q/ds =\partial
  H/\partial \delta p}$, ${d\delta p/ds =-\partial H/\partial \delta
  q}$ to Eq.~(\ref{eq:Hsqsmall}) yields the paraxial ray tracing
equations, a system of ordinary differential
equations~\cite{Bos09,Cerveny87,Cerveny01,Jensen11,Jobert87,Keller83,Nair91,Porter87,Shen75}
\begin{dbleqnarray} {d\delta q\over ds} &=&c_s\,\delta p, &\delta
q(0)&=&0, \label{eq:delqde} \\ {d\delta p\over ds} &=&-{c_{\rm
    nn}\over c_s^2}\,\delta q, &\delta
p(0)&=&c_0^{-1}\,\delta\theta_0,\label{eq:delpde}\end{dbleqnarray} the
initial conditions arising from the facts that the spread of rays
separated by a small angle~$\delta\theta_0$ is zero at~${s=0}$, and
the differential time equals ${\delta T=c_0^{-1}\,\delta
  q\,\delta\theta_0}$.

\subsection{\label{sec:paraxial-ray-intrinsic}Paraxial Ray Equations:
  Sturm-Liouville Form}

Before explaining the full significance of the complete second-order
terms in the paraxial ray equations in the next section, it will be
helpful to express Eqs.\ (\ref{eq:delqde})--(\ref{eq:delpde}) as a
single second-order differential equation using the natural or
intrinsic parameters of the problem. Consistent with much of existing
literature on paraxial rays, rays are parameterize using the
arclength~$s$.  However, physically rays do not minimize arclength,
but travel-time~$t$, and therefore the paraxial ray's underlying
physical and geometric properties will be revealed by using the
intrinsic parameterization $dt=c^{-1}\,ds$. Also the variables $\delta
q$ and $\delta p=\partial T/\partial\delta q$ are expressed using
arclength or distance; the corresponding intrinsic variables arise by
converting distance~$\delta q$ to travel-time $\delta\tilde q$ and
slowness~$\delta p$ to the dimensionless~$\delta\tilde p$ using the
sound speed~$c$: \begin{equation} \delta\tilde q\buildrel{\rm
    def}\over=c^{-1}\,\delta q, \quad\hbox{and}\quad \delta\tilde
  p\buildrel{\rm def}\over=c\,\delta
  p. \label{eq:tildeqpdef} \end{equation} Note that $\delta\tilde q$
is the travel-time between nearby rays and $\delta\tilde p=\partial
T/\partial\delta\tilde q$ is the travel-time derivative along a
straight-line path normal. The geometric significance of these
intrinsic coordinates is made clear by the following new theorem:

\begin{theorem}\label{thm:pqjacobi}Let $\delta q$ and $\delta p$ satisfy Eqs.\
(\ref{eq:delqde})--(\ref{eq:delpde}). Then these extrinsic first-order
  paraxial ray equations are equivalent to the intrinsic second-order
  paraxial ray
  equation \begin{equation}\label{eq:paraxial-sturm-liouville}
    \delta\ddot{\tilde q} +\bigl(cc''-(c')^2\bigr)\,\delta\tilde
    q=0,\end{equation} where $\delta\tilde q=c^{-1}\,\delta q$,
  $\delta\tilde p=c\,\delta p$, $\delta\ddot{\tilde
    q}=(d^2/dt^2)\delta\tilde q$, and the coefficient ${cc''-(c')^2}$
  is the Gaussian curvature of the propagation medium w.r.t.\ to the
  Fermat metric of Eq.~(\ref{eq:pathtime}).\end{theorem}

The proof involves parameterizing by travel-time~$t$ and expressing
the Hamilton form of the paraxial ray equations of
Eqs.\ (\ref{eq:delqde})--(\ref{eq:delpde}) as the coupled first-order
system, \begin{debpeqnarray} \delta\dot{\tilde q}
&=&-c'(dz/ds)\,\delta\tilde q +\delta\tilde p &\delta\tilde q(0)&=&0,
\label{eq:tildeqde} \\ \delta\dot{\tilde p} &=&-cc_{\rm nn}\,\delta\tilde q
+c'(dz/ds)\,\delta\tilde p &\delta\tilde p(0)&=&\delta\theta_0.\label{eq:tildepde}
\end{debpeqnarray} Taking another derivative of the first equation
yields ${\delta\ddot{\tilde q} +K\,\delta\tilde q=0}$ with
$K={cc''-(c')^2}$. The theorem is thus proven by demonstrating that
this expression for~$K$ is precisely the Gaussian curvature
w.r.t.\ the Fermat metric of Eq.~(\ref{eq:pathtime}). Given an
arbitrary metric $g$ on a $2$-dimensional manifold with coefficients
${g_{ij}=g(\Xbmit_i,\Xbmit_j)}$ w.r.t.\ a basis $\{\,\Xbmit_i\,\}$,
the Gaussian curvature is determined~\cite{Rund59,Spivak99} by the expression
$K =R_{1212}/({g_{11}g_{22}-g_{12}^2})$ in which
$R_{ijkl}=\sum_ng_{in}R^n{}_{jkl}$, $R^i{}_{jkl}
=(\partial\Gamma^i_{lj}/\partial x^k) -(\partial\Gamma^i_{kj}/\partial
x^l) +\sum\nolimits_n({\Gamma^n_{lj}\Gamma^i_{kn}
  -\Gamma^n_{kj}\Gamma^i_{ln}})$ are the coefficients of the
Riemannian curvature tensor~$\Rbmit$, and the Christoffel symbols (of
the second kind) $\Gamma^k_{ij} ={1\over2}\sum\nolimits_l
g^{kl}\bigl((\partial g_{il}/\partial x^j) +(\partial g_{jl}/\partial x^i)
-(\partial g_{ij}/\partial x^l)\bigr)$ appear in the quadratic coefficients
in the ray (geodesic) equations \begin{equation}\textstyle \ddot x^k
  +\sum\nolimits_{ij}\Gamma^k_{ij}\dot x^i\dot
  x^j=0\label{eq:geodesicequation}\end{equation} with
${(g^{kl})=(g_{ij})^{-1}}$ representing the matrix inverse of the
matrix of metric coefficients.~\cite{Helgason78,Rund59,Spivak99} The
Christoffel symbols expressed as quadratic forms are
\begin{equation} (\Gamma^1_{ij})
=\begin{pmatrix}0&-c'/c\\-c'/c&0\end{pmatrix},\quad (\Gamma^2_{ij})
=\begin{pmatrix}c'/c&0\\0&-c'/c\end{pmatrix}, \label{eq:FermatChristoffel}\end{equation}
can be read directly from the ray equations of Eq.~(\ref{eq:rayrz2});
these along with the matrix
$(g_{ij})=c^{-2}\Bigl({1\atop0}\;{0\atop1}\Bigr)$ can be used directly
with the preceding equations to compute the Gaussian curvature
${K=cc''-(c')^2}$.  As will be proved,
Eq.~(\ref{eq:paraxial-sturm-liouville}) is precisely Jacobi's
equation; therefore, the Hamiltonian form of the paraxial ray
equations (\ref{eq:delqde})--(\ref{eq:delpde}) is mathematically and
physically equivalent to the intrinsic Jacobi's equation. This
equivalence between \Cerveny's formulation of Gaussian beams and
Jacobi's equation is a new result.  The functional form of the
Gaussian curvature will be used to derive the distance between
convergence zones and ``model spaces'' of constant curvature for the
SSP.

\section{\label{sec:geometric-spreading}Intrinsic Geometry of Gaussian
Beams}

\subsection{\label{sec:geometric-amplitude}Transverse Amplitude of a Gaussian
Beam}

This section contains an analysis of Gaussian beams that use Jacobi's
equation to quantify their amplitude and phase.  All the results
obtained so far have been for rays and infinitesimal variations around
them. In contrast, Gaussian beams have real, finite physical width
that must be accounted for in the plane transverse to the ray. The
physical width has different implications for the Gaussian beam's
amplitude and phase. The amplitude is determined by both the geometric
spreading loss along the ray and the initial amplitude distribution in
the transverse plane.  The phase is determined by the differential
change in travel-time along the transverse plane.  Mathematically, the
scaled normal vector $\delta q\,\Nb$ represents the
infinitesimal change in ray position as a function of the take-off
angle for a fixed path length~$s$. Gauss's lemma implies that this
travel-time is fixed to first-order, but says nothing about the
second-order terms, i.e.\ $\xb(s;\theta_0) +\delta q\Nb
=\xb(s;{\theta_0 +\delta\theta_0}) +O(\delta\theta_0^2)$
for~${\delta\theta_0>0}$.  This ray-front curvature effect is seen in
Euclidean space by holding a ruler next to a circle or cylinder and
observing the size of gap---which for a circle of radius~$R$ has a gap
size of ${R\sec \delta\theta-R}
={(R/2)\delta\theta^2+O(\delta\theta^4)}$.  The second-order
travel-time difference is often physically significant because it is
of the order of several wavelengths at a broad range of frequencies
$f_0$ and therefore does contribute to the Gaussian beam's phase~$2\pi
f_0t$. Because the travel-time difference in the transverse plane is
quadratic, the phase of Gaussian beam necessarily has a Gaussian
distribution. Furthermore, these second-order affects are already
accounted for in the Riemannian curvature terms determining the
Gaussian beam's amplitude, so the Gaussian beam's amplitude in the
transverse plane is determined by the effect of geometric spreading on
initial conditions. In this subsection we will quantify the impact of
the initial conditions on the amplitude and the second-order
travel-time differences on the phase.  Both quantities are determined
by the ray distance $\delta q$, which will now be formalized for both
extrinsic and intrinsic geometric spreading.

\begin{definition}\label{defn:geometric-spreading}{\rm(Geometric
    spreading)}\quad Let $\xb(s;\theta_0) =\bigl(r(s;\theta_0),
  z(s;\theta_0)\bigr)^\T$ be a ray parameterized by arclength~$s$
  with initial conditions $\xb(0;\theta_0)=(0,z_0)^\T$ and
  $(d/ds)\xb(0;\theta_0) =(\cos\theta_0,-\sin\theta_0)^\T$ in a
  horizontally stratified media with SSP $c(z)$. The\/ {\em extrinsic
    geometric spreading}~$q(s)$ along the ray caused by a
  change in elevation angle $\theta_0$ is defined to be
  \begin{equation} q(s)
  =\|(\partial\xb(s; \theta_0) /\partial\theta_0)_s
  \|_2,\label{eq:geometric-spreading}
  \end{equation} where $\|{\cdot}\|_2$ denotes the standard
  $2$-norm and the standard notation
  $(\partial/\partial\theta_0)_s$ means partial differentiation
  w.r.t.\ $\theta_0$ while holding arclength~$s$ constant.
\end{definition}

\begin{theorem}{\rm(Geometric spreading for horizontally stratified
    media~\cite{Jensen11})}\quad \label{thm:geometric-spreading}Let
  $\xb(s)=\bigl(r(s;\theta_0),z(s;\theta_0)\bigr)^\T$ be a ray in a
  horizontally stratified medium with a twice-differentiable SSP
  $c(z)$ parameterized by arclength~$s$ and with initial elevation
  angle~$\theta_0$.  The extrinsic geometric spreading~$q$ and the
  corresponding canonical variable $p=\partial T/\partial q$ along the
  ray are determined by \begin{equation} q
    =\lim_{\delta\theta_0\to0}\delta q/\delta\theta_0,\qquad p
    =\lim_{\delta\theta_0\to0}\delta
    p/\delta\theta_0,\label{eq:qpgsintrspr}\end{equation} and satisfy
  the Hamilton equations \begin{dbleqnarray} dq/ds &=&cp,
    &q(0)&=&0, \label{eq:qde} \\ dp/ds &=&-(c_{\rm nn}/c^2)q,
    &p(0)&=&c_0^{-1}.\label{eq:pde}\end{dbleqnarray}\end{theorem}

The proof is a trivial application of the definition of the small or
infinitesimal difference $\delta q$ defined in
Eq.~(\ref{eq:qdef}). The theorem follows immediately by dividing
Eqs.\ (\ref{eq:delqde})--(\ref{eq:delpde}) by~$\delta\theta_0$ and
taking the limit.  Thus we have a complete description of the
extrinsic geometric spreading for Gaussian beams, including the
complete second-order terms introduced in Eq.~(\ref{eq:delpde}). To
appreciate the geometric significance of the spread, we will define
the {\em intrinsic\/} geometric spreading and equate this concept with
Jacobi's equation first encountered in Theorem~\ref{thm:pqjacobi},
expressed in this new theorem:

\begin{definition}{\rm(Intrinsic geometric
    spreading for horizontally stratified
    media)}\quad \label{defn:intrinsic-geometric-spreading}Let
  ${\bigl(r(t;\theta_0),z(t;\theta_0)\bigr)}$ be a ray in a horizontally stratified
  medium with a twice-differentiable SSP $c(z)$ as in
  Theorem~\ref{thm:geometric-spreading}, but parameterized by path
  time~$t$. The\/ {\em intrinsic geometric
    spreading}~$\intrgeomspr(t)$ along the ray is defined to be
  \begin{equation} \intrgeomspr(t)
    =\lim_{\delta\theta_0\to0}{\delta\tilde
      q\over\delta\theta_0}=c^{-1}q(t)=\tilde
    q(t),\label{eq:intrgeomsprdef}
  \end{equation} where $\delta\tilde q$ is the travel-time between
  nearby rays separated by~$\delta\theta_0$ at their initial point
  defined in Eq.~(\ref{eq:tildeqpdef}), and $q(t)$ is the extrinsic
  geometric spreading parameterized by~time.\/ {\em [Intrinsic
      geometric spreading is denoted by the Arabic letter `ayn,
      (`$\intrgeomspr$', pronounced like the end of `nine' spoken with
      a strong Australian accent) in recognition of the mathematician
      Ibn~Sahl who discovered Snell's law of refraction around
      984. Ibn~Sahl used the symbol~$\intrgeomspr$ to denote the
      center of a lens~\cite{Rashed90}.]}
\end{definition}


The intrinsic geometric spreading is a consequence of an obvious
corollary to Theorem~\ref{thm:geometric-spreading} arising from
Eqs.\ (\ref{eq:tildeqde})--(\ref{eq:tildepde}) for the intrinsic
variables \begin{equation} \tilde q =c^{-1}q
  =\lim_{\delta\theta_0\to0}\delta\tilde q/\delta\theta_0
  =\intrgeomspr, \quad\hbox{and}\quad\tilde p =cp
  =\lim_{\delta\theta_0\to0}\delta\tilde p/\delta\theta_0
  =\dotintrgeomspr+c'\sin\theta\intrgeomspr.\label{eq:ptilde}\end{equation}
Note that the path time derivative ${\tilde p=\partial
  T/\partial\tilde q}$ w.r.t.\ distances along the extrinsic path
normal~$\Nb$ is {\em not\/} exactly equal to the derivative
$\dotintrgeomspr=(d/dt)\intrgeomspr$ of the intrinsic geometric
spreading; the geometric reason for this difference will be explained
in section~\ref{sec:gb-ampphase}. We are now ready to introduce
Jacobi's equation and thereby compute the ray's amplitude determined
by the geometric spreading, as well a Gaussian beam's phase in the
transverse plane perpendicular to the ray's tangent.

\begin{theorem}{\rm(Intrinsic geometric spreading for horizontally stratified
    media)}\quad \label{thm:jacobi-equation2}Let
  ${\bigl(r(t;\theta_0),z(t;\theta_0)\bigr)^\T}$ be a ray in a
  horizontally stratified medium with a twice-differentiable SSP $c(z)$
  parameterized by path time~$t$ and with initial elevation
  angle~$\theta_0$.  The intrinsic geometric
  spreading~$\intrgeomspr(t)$ along the ray satisfies the
  Sturm-Liouville equation \begin{equation} \ddot\intrgeomspr
    +K(t)\intrgeomspr =0; \quad
    \intrgeomspr(0)=0,\ \dotintrgeomspr(0)=1,\label{eq:Jacobi2d}
  \end{equation} where
  \begin{equation} K =cc''-(c')^2
  \label{eq:acousticGaussiancurvature} \end{equation} is the acoustic Gaussian
curvature of the Fermat metric $g(\dot\xb,\dot\xb) =c^{-2}(z)(\dot
r^2+\dot z^2)$ of Eq.~(\ref{eq:pathtime}).
\end{theorem}

Eq.~(\ref{eq:Jacobi2d}) is simply the intrinsic Jacobi's equation in
two dimensions encountered above in
Eq.~(\ref{eq:paraxial-sturm-liouville}).  The theorem follows
immediately by dividing Eq.~(\ref{eq:paraxial-sturm-liouville})
by~$\delta\theta_0$ and taking the limit.  In general, Jacobi's
equation for an arbitrary Riemannian manifold is, \begin{equation}
  \nabla_\Tbmit^2\Vbmit +\Rbmit(\Vbmit,\Tbmit)\Tbmit
  =\zerob,\label{eq:genjacobi} \end{equation} where
$\nabla_\Tbmit^2\Vbmit$ is the second covariant derivative along the
ray's tangent vector $\Tbmit =(\dot r,\dot z)^\T$ of the variation
vector $\Vbmit =\delta\xb/\delta\theta_0 =\intrgeomspr\Ybmit$ along
the unit vector ${\Ybmit=-(\dot z,-\dot r)^\T}$, and the Riemannian
curvature tensor equals \begin{equation} \Rbmit(\Xbmit,\Ybmit)\Zbmit
  =\nabla_{\!\Xbmit}\nabla_\Ybmit\Zbmit
  -\nabla_\Ybmit\nabla_{\!\Xbmit}\Zbmit
  -\nabla_{[\Xbmit,\Ybmit]}\Zbmit,
\label{eq:Rcurvedef} \end{equation} 
where ${[\Xbmit,\Ybmit]} ={\Xbmit\Ybmit -\Ybmit\!\Xbmit}$ is the Lie
bracket.~\cite{Cheeger75,Helgason78,Spivak99} Note that the
orthonormal frame $(\Tbmit,\>\Ybmit)$ is parallel along the ray, i.e.\
${\nabla_\Tbmit\Tbmit=\zerob}$ [Eq.~(\ref{eq:rayrz2})]
and~${\nabla_\Tbmit\Ybmit=\zerob}$ so that $\nabla_\Tbmit\Vbmit
=\dotintrgeomspr\Ybmit +\intrgeomspr\nabla_\Tbmit\Ybmit
=\dotintrgeomspr\Ybmit$. Furthermore, ${[\Tbmit,\Vbmit]\equiv\zerob}$
because the vector fields $\Tbmit$ and~$\Vbmit$ are defined as
independent partial derivatives of the ray~$\xb(t;\theta_0)$;
therefore, ${\nabla_\Tbmit\Vbmit=\nabla_\Vbmit\Tbmit}$ by the property
${\nabla_{\!\Xbmit}\Ybmit -\nabla_\Ybmit\Xbmit =[\Xbmit,\Ybmit]}$ of
covariant differentiation.  Taking another covariant derivative
w.r.t.\ $\Tbmit$ and computing an inner product with~$\Ybmit$ along
with the definition for Gaussian
curvature~\cite{Cheeger75,Helgason78,Spivak99} \begin{equation} K
  =g\bigl(\Rbmit(\Ybmit,\Tbmit)\Tbmit,\Ybmit\bigr)
  /\bigl(g(\Tbmit,\Tbmit)g(\Ybmit,\Ybmit)
  -g(\Tbmit,\Ybmit)^2\bigr)\label{eq:SectionalK}
\end{equation} yields Jacobi's equation of Eq.~(\ref{eq:Jacobi2d}).

Jacobi's equation provides a physically appealing description of
Gaussian beams; indeed, interpretations of Jacobi's equation and
Gaussian curvature provide the following new insights into the problem
of ray acoustics. At depths where the SSP $c(z)$ is concave, i.e.\ a
sound duct below which rays refract upwards and above which they
refract down, $cc''>0$, $c'\approx0$, and the acoustic Gaussian
curvature is positive, yielding a sinusoidal behavior with
wavelength---the distance between caustics---of about $\pi
cK^{-{1\over2}}$ for the geometric spreading, as is expected and
encountered with caustics in sound duct (Figs.\ \ref{fig:munk-ssp},
\ref{fig:cosh-ssp}).  Note well that caustics, defined to be points
where the geometric spreading vanishes (these are called ``conjugate
points'' in the mathematical literature), will occur at integer
multiples of the range \begin{equation}
  {\hbox{half-wavelength}\atop\hbox{distance}}\approx \pi
  cK^{-1/2}\approx
  \pi(c/c'')^{1/2}.\label{eq:half-wavelength}\end{equation} (the last
approximation valid when the SSP's second derivative dominates). At
depths where the SSP $c(z)$ is convex, i.e.\ divergent zones below
which rays refract downwards and above which they refract up, or at
depths with a linear SSP, $cc''\le0$, $(c')^2>0$, and the acoustic
Gaussian curvature is negative, yielding an exponentially growing
solution to geometric spreading with length constant of about
$(-K)^{-1/2}$, and therefore a large spreading loss.  Constant SSPs
imply that the acoustic Gaussian curvature is zero, yielding geometric
spreading that simply grows linearly with the path length, as
expected.  At interfaces where the SSP or its first derivative are
discontinuous, such as the bottom or surface where the method of
images is used to model reflections, Eq.~(\ref{eq:Jacobi2d}) can be
integrated using the standard modifications, provided below,
necessitated by the Dirac delta function. These observations will all
be formalized in section~\ref{sec:linear-const-curv-ssp} below.  It
also shown below that the known extrinsic formulae for the geometric
spreading loss based on Snell's law and variants of the extrinsic
Frenet-Serret formulae~\cite{Cerveny87,Cerveny01,Porter87,Jensen11}
satisfy Jacobi's equation.

Eq.~(\ref{eq:Jacobi2d}) may be proven directly via the second
variation, but this direct approach, which involves a long, messy
computation that provides almost no geometric insight, motivates
introduction of the cleaner and simpler modern covariant approach.
Nevertheless, direct computation establishes one important result for
readers without backgrounds in covariant differentiation and
Riemannian geometry, so we will provide a sketch in this paragraph.
Gauss's lemma assures us that nearby rays are perpendicular to a given
ray's tangent vector $(\dot r,\dot
z)^\T$;~\cite{Gelfand63,Helgason78,Spivak99} therefore, nearby rays
all take the form $\delta\xb =\bigl(\intrgeomspr(t)\dot
z\,\delta\theta_0, -\intrgeomspr(t)\dot r\,\delta\theta_0\bigr)^\T$
for the geometric spreading function~$\intrgeomspr(t)$ along the
ray. It can be shown using the Hessian matrix of the Fermat metric of
Eq.~(\ref{eq:pathtime}) w.r.t.\ $(\xb,\dot\xb)$, and application
of the Christoffel Eq.~(\ref{eq:rayrz2}), that the second
variation~\cite{Gelfand63} of the travel-time of
Eq.~(\ref{eq:1stvarfor}) equals \begin{multline} \delta^2 L[\delta\xb]
  =L[\xb+\delta\xb]-L[\xb]-\delta L[\delta\xb] -o(\|\delta\xb\|^2)
  \\=\delta\theta_0^2\int_0^{t_{\rm end}} \biggl(\Bigl(
  \bigl(-c^{-2}(cc''-(c')^2\bigr) +(c'/c)\ddot z\Bigr)\intrgeomspr^2
  +\dotintrgeomspr^2 +2(c'/c)\dot
  z\intrgeomspr\dotintrgeomspr\biggr)\,dt.
\label{eq:2dvarfor} \end{multline} Because these second order terms
necessarily satisfy the Euler-Lagrange equation for nearby extremals
$\xb+\delta\xb$, the function $\delta\xb$ must satisfy the
Euler-Lagrange equation for the second variation, known as Jacobi's
equation. Applying the Euler-Lagrange equation for the unknown
function $\intrgeomspr(t)$ to the function appearing in the second
variation of Eq.~(\ref{eq:2dvarfor}) yields Jacobi's equation of
Eq.~(\ref{eq:Jacobi2d}). This establishes
Theorem~\ref{thm:jacobi-equation2}, but it does not provide any
immediate connection to the problem's geometry or intrinsic curvature,
nor does it suggest a method to generalize the result to three (or
more) dimensions.  These connections will now be established within
the broadly general framework of covariant differentiation.

\subsection{\label{sec:geometric-spread-prop}Properties of Geometric Spreading}

In this section we will review, in the context of the results
presented above, some simple properties about geometric spreading.
First, the geometric spreading $q(t)$ quantifies the distance
of nearby rays from the ray $\xb(t)$, parameterized by its initial
elevation angle $\theta_0$. Second, in the trivial case of a constant
SSP, geometric spreading simply equals range, ${q(t) =ct}$
(constant SSP).  Third, and crucially for understanding the results
presented in this paper, the direction of geometric spreading is
perpendicular to the ray's (extrinsic) tangent vector,
$\Tb=d\xb/ds$, and is therefore also perpendicular to the ray's
intrinsic tangent vector $\dot\xb =(d\xb/ds)(ds/dt) =c(z)\Tb$. This
orthogonality property, called Gauss's lemma because it is among of
the first key observations by Gauss in his establishment of the
intrinsic theory of surfaces, is also a straightforward consequence of
Euler's first variation of the functional $L[\xb(t)] =\int_0^{t_{\rm
    end}} F(t,\xb,\dot\xb)\,dt$ [as in Eq.~(\ref{eq:pathtime})] along
an extremal $\xb(t)$ with only the initial endpoint fixed.  The first
variation formula for an arbitrary functional variation $\delta\xb(t)$
such that ${\delta\xb(0)=0}$ is
\begin{multline} \delta L[\delta\xb] =L[\xb+\delta\xb]-L[\xb] \\
 =\int_0^{t_{\rm end}}
\left({\partial F\over\partial\xb}-{d\over dt} {\partial
    F\over\partial\dot\xb}\right)\,\delta\xb\,dt
+\left.{\partial
    F\over\partial\dot\xb}\,\delta\xb\right|_0^{t_{\rm end}}
 = (\partial F/\partial\dot\xb)\,\delta\xb(t_{\rm
  end})=0, \label{eq:1stvarfor}
\end{multline} where the integral vanishes because $\xb$ is assumed to be an
extremal and satisfies the Euler-Lagrange equation, and
${\delta\xb(0)=0}$ because the starting point is fixed.  The first
variation must vanish necessarily for arbitrary functions
$\delta\xb(t)$; therefore ${(\partial
  F/\partial\dot\xb)\,\delta\xb(t_{\rm end})=0}$.  In our specific
case of minimum travel-time rays, $F(t,\xb,\dot\xb)=c^{-1}(z)(\dot
r^2+\dot z^2)^{1/2}$ and
$\delta\xb=(\partial\xb/\partial\theta_0)_t\,\delta\theta_0$ and
Gauss's lemma follows immediately from Eq.~(\ref{eq:1stvarfor}) for
the case of acoustic rays:
\begin{equation} c^{-2}(z)\left(\left({dr\over
        dt}\right)_{\theta_0} \left({\partial r\over
    \partial\theta_0}\right)_{t} +\left({dz\over dt}\right)_{\theta_0}
  \left({\partial z\over \partial\theta_0}\right)_{t}\right)
  =0,\label{eq:Gausslemma}\end{equation} Note that this basic argument
holds for all extremal paths, whether or not they are minimizing or
pass through a conjugate point a.k.a.\ caustic.  Consequently from
Gauss's lemma, a fourth property is that the geometric spreading
depends on both $\dot r$- and~$\dot z$-components of the ray's tangent
vector.  Fifth, there is a geometric spreading term in the
range-azimuthal $(r,\phi)$-plane, which, by symmetry, is trivially
$\partial q(t)/\partial\phi_0 =r(t)$.  This intuitively obvious
fact will be proven as an illustration of the covariant formalism
introduced below.  Sixth, and finally, geometric spreading as defined
here is physically extrinsic (relative to the metric of travel-times
between points) because pressure transducers integrate power over
physical area, which is why the Euclidean or $2$-norm appears in
Definition~\ref{defn:geometric-spreading}. Nonetheless, a simple
change of scale [multiplication by~$1/c(z)$], converts the problem of
geometric spreading to an intrinsic framework that depends only upon
the travel-time between points, i.e.\ only on rays.

The transmission loss along a ray tube caused by geometric spreading
is given by the infinitesimal area at a standard reference range
($r_{\rm ref}\buildrel{\rm def}\over= 1\,$m) divided by the
infinitesimal area of the ray tube at an arbitrary distance~$t$, both
scaled by the inverse sound speed along the
ray~\cite{Jensen11}: \begin{equation} \hbox{TL (geom.)}
  =\lim_{\delta\phi,\delta\theta\to0} {c_sr_{\rm ref}^2\cos\theta_0\,
    \delta\theta\, \delta\phi\over c_0rq(t)\, \delta\theta\,
    \delta\phi} ={c_s\cos\theta_0\over c_0rq}\quad\hbox{(re $1\,{\rm
      m}^2$).}\label{eq:TLg}
\end{equation} When the density also varies with depth, Newton's
    second law must be included: \begin{equation} \hbox{TL (geom.)}
      ={\rho_sc_s\cos\theta_0\over \rho_0c_0 rq}\quad\hbox{(re
        $1\,{\rm m}^2$).}\label{eq:TLgp}
\end{equation}

\begin{figure}[t]
\begin{multicols}{2}
\begin{figure}[H]
\centerline{\includegraphics[width=\columnwidth]{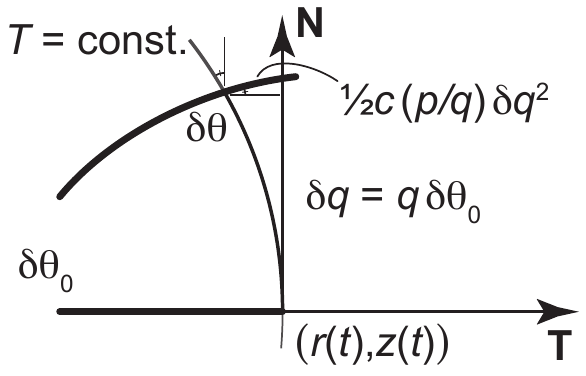}}
\caption{Phase of the Gaussian beam along the ray's extrinsic
  normal~$\Nb=d\Tb/ds$. The differential time~$\delta t_{\rm e}$ a
  distance~$\delta\eta$ along the normal is given by the quadratic
  $\delta t_{\rm e} =\texthalf(p/q)\,\delta\eta^2 =\texthalf
  c^{-2}(\dotintrgeomspr/\intrgeomspr+c'\sin\theta)\,\delta\eta^2$. All
  distances in the figure represent extrinsic Euclidean
  distances.\label{fig:beam-phase}}
\end{figure}
\begin{figure}[H]
\centerline{\includegraphics[width=\columnwidth]{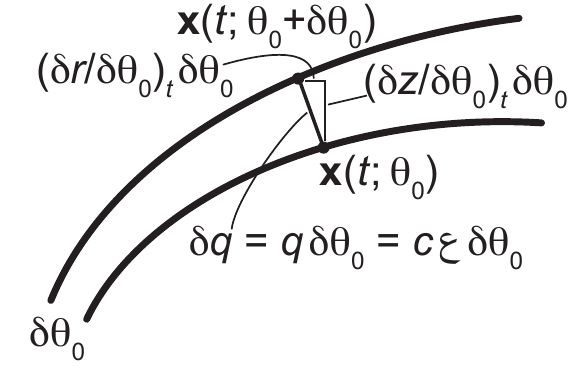}}
\caption{Geometric spreading expressed using ray-centered and
  extrinsic range and depth coordinates. All distances in the figure
  represent extrinsic Euclidean distances.\label{fig:snellgeomspread}}
\end{figure}
\end{multicols}
\vspace{-12pt}
\end{figure}

\subsection{\label{sec:geometric-phase}Transverse Phase of a Gaussian Beam}

The differential time $\delta t_{\rm e}$ perpendicular to the ray
(extrinsic) determines the differential phase~$e^{j\,\delta\varphi}$
of the Gaussian beam, where ${\delta\varphi=2\pi f_c\,\delta t_{\rm
    e}}$, $f_c$ is the center or carrier frequency of the sound
propagating along the ray, and ${j=\sqrt{-1}}$. In terms of the
extrinsic geometric spreading variables $q$ and~${p=\partial
  T/\partial q}$, the differential time a distance~$\delta\eta$ along
the ray normal at~${\bigl(r(s),z(s)\bigr)^\T}$ is given by the
formulae~\cite{Cerveny01,Jensen11} \begin{equation} \delta t_{\rm
    e}={1\over2}{p\over q}\,\delta\eta^2 ={1\over
    2}\left({\dotintrgeomspr\over \intrgeomspr}
  +c'\sin\theta\right)\,\delta\mu^2, \label{eq:extdifftime}\end{equation}
where ${\delta\mu=c^{-1}\,\delta\eta}$.  This fact is illustrated in
Fig.~\ref{fig:beam-phase}.  To compare this extrinsic formula with the
intrinsic results developed in section~\ref{sec:gb-ampphase}, it is
worthwhile to give a brief geometric proof of
Eq.~(\ref{eq:extdifftime}). By inspection, or formally by Gauss's
lemma of Eq.~(\ref{eq:Gausslemma}) the differential path time along
the ray normal is a quadratic function of distance up to second order,
i.e.\ \begin{equation} \delta t_{\rm e}=\texthalf Q_{\rm
    e}\,\delta\eta^2\label{eq:Qdef}\end{equation} for some quadratic
coefficient~$Q_{\rm e}$ to be determined. As seen in
Fig.~\ref{fig:beam-phase}, the slope of this quadratic function
at~$\delta\eta$ equals $\delta\theta$, the differential change in
angle: \begin{equation} \delta\theta =cd(\delta t_{\rm
    e})/d(\delta\eta)|_{\delta\eta=\delta q=q\,\delta\theta_0}
  =cQ_{\rm e}\,\delta q=cqQ_{\rm
    e}\,\delta\theta_0.\label{eq:dtdmu}\end{equation} However,
$\delta\theta$ may be computed directly via an inner product along
with Eq.~(\ref{eq:qde}), \begin{equation} \delta\theta
  =\langle\Nb,(d/ds)\xb(s;\theta_0+\delta\theta_0)\rangle=
  \bigl\langle\Nb,(d/ds)\bigl(\xb(s;\theta_0)
  +q\,\delta\theta_0\,\Nb\bigr)\bigr\rangle
  =cp\,\delta\theta_0 \label{eq:Ndxdsip}\end{equation} Equating
$\delta\theta$ from Eqs.\ (\ref{eq:dtdmu}) and~(\ref{eq:Ndxdsip})
yields ${Q_{\rm e}=p/q}$, from which Eq.~(\ref{eq:Qdef}) becomes the
first equality of Eq.~(\ref{eq:extdifftime}). Applying
Eq.~(\ref{eq:tildeqde}) to Eq.~(\ref{eq:extdifftime})
yields \begin{equation} c^2{p\over q} ={\dotintrgeomspr\over
    \intrgeomspr} +c'\sin\theta \label{eq:c2pq}\end{equation} and the
second equality.

Summarizing the results of the previous sections, the transverse
amplitude and phase of a Gaussian beam emanating a
distance~$\delta\eta$ normal to the ray and with initial source
level~$\hbox{SL}$ and azimuth end elevation angle fan widths
$\delta\phi_0$ and~$\delta\theta_0$ is given by \begin{eqnarray}
  P(s;\theta_0;\delta\mu)
  &=&\left[{\hbox{SL}{\cdot}c_s\cos\theta_0\over
        c_0rq}\right]^{{1\over2}}\exp\left[j2\pi
      f_c\left(t(s)+{p\over2q}\,\delta\eta^2\right)\right]
  \label{eq:extrayampphase}\\ &&\left[{\hbox{SL}{\cdot}a\over
        r\intrgeomspr}\right]^{{1\over2}}\exp\left[j2\pi
      f_c\left(t+{1\over2}\left({\dotintrgeomspr\over\intrgeomspr}+c'\sin\theta\right)\,\delta\mu^2\right)\right].
  \label{eq:intrayampphase}\end{eqnarray} 
Note that Gaussian beam is an appropriate name for rays with this
functional form. A Gaussian distribution of amplitude is typically
introduced through the initial conditions for the Gaussian beam.

\subsection{\label{sec:linear-const-curv-ssp}Munk, Linear, and SSPs with Constant Curvature}

\subsubsection{\label{sec:munk-ssp}Munk Profile}

The distance between convergence zones\slash cycle distances for the
Munk profile $c(z) =c_0\bigl(1+\epsilon(\bar z+e^{-\bar z}-1)\bigr)$
with ${\bar z=(z-z_0)/W}$ can be computed directly via
Eq.~(\ref{eq:half-wavelength}). At ${z=z_0}$, the Gaussian curvature
$K=cc''-(c')^2 =\epsilon c_0^2/W^2$, proving the new theorem:

\begin{theorem}\label{thm:munkdist}The cycle distance of the Munk profile
approximately equals \begin{equation} {\hbox{Munk
      profile}\atop\hbox{CZ distance}}\approx
  \pi\epsilon^{-1/2}W.\end{equation}
\end{theorem}

E.g.\ for the nominal parameters ${\epsilon=0.00737}$ and
${W=650\,}$m, this yields a distance of~$23.8\,$km
(Fig.~\ref{fig:munk-ssp}).

\subsubsection{\label{sec:linear-ssp}Linear SSPs}

The ideas developed to this point are nicely illustrated using the
example of Gaussian beams in medium with linear SSP ${c(z)=c_0+\gamma
  z}$, in which case rays are simply circles
in~$(r,z)$-space:~\cite{Jensen11} \begin{eqnarray} r(\theta)
  &=&z_{0\gamma}\sec\theta_0\bigl(\sin\theta_0
  +\sin(\theta-\theta_0)\bigr),\\ z(\theta)
  &=&-c_0\gamma^{-1}+z_{0\gamma}\sec\theta_0
  \cos(\theta-\theta_0), \end{eqnarray} where the ray emanates
from~$(0,z_0)$ with initial elevation angle~$\theta_0$, and
${z_{0\gamma}=z_0+c_0\gamma^{-1}}$. Arclength is given by
${s=z_{0\gamma}\theta}$, and eigenrays to~$(r,z)$ are determined by
the angles \begin{eqnarray} \theta_0 &=&2\atantwoactual
  \Bigl(2rz_{0\gamma}-\bigl(r^2 + (z_{0\gamma}-z)^2\bigr)\bigl(r^2 +
  (z_{0\gamma}+z)^2\bigr),
  z_{0\gamma}^2-z^2-r^2\Bigr)^{1\over2},\\ \theta &=&
  \atantwoactual(r-z_{0\gamma}\tan\theta_0,z+c_0\gamma^{-1})+\theta_0,\end{eqnarray}
with travel-time \begin{equation} t ={1\over\gamma}
  \log\left({\tan\texthalf \sin^{-1}\cos\theta_0\over \tan\texthalf
    \sin^{-1}\left({c_0+\gamma z\over c_0+\gamma
      z_0}\cos\theta_0\right)}\right),\end{equation} assuming no
caustics along the ray. Because ${c_{\rm nn}\equiv0}$, the Hamilton
equations (\ref{eq:qde})--(\ref{eq:pde}) are easily solved in closed
form, \begin{equation} q = z_{0\gamma}\sec^2\theta_0
  \Bigl(\bigl(1-a^2(c_0+\gamma z)^2\bigr)^{1/2} -|\sin\theta_0|\Bigr),
  \quad\hbox{and}\quad p ={1\over c_0+\gamma
    z_0}. \label{eq:linear-ssp-q}\end{equation} Likewise, because the
Gaussian curvature ${K\equiv -\gamma^2}$ is constant (the propagation
medium is a space of constant negative curvature, i.e.\ it is a
hyperbolic space), Jacobi's equation (\ref{eq:Jacobi2d}) has the
closed-form solution $\intrgeomspr(t) =\gamma^{-1}\sinh \gamma t$, or
${\intrgeomspr(t) \approx t+{1\over6}\gamma^2t^3}$
for~${t\lessapprox\gamma^{-1}}$, yielding $q=c\gamma^{-1}\sinh\gamma t
\approx {ct(1+{1\over6}\gamma^2t^2)}$, a much simpler expression than
Eq.~(\ref{eq:linear-ssp-q}). The expression for the relative delay is
given by ${p/q\approx c_1^{-1}c^{-1}t^{-1}(1-{1\over6}\gamma^2t^2)}$,
a new result.

\subsubsection{\label{sec:const-curv-ssp}SSPs with Constant Curvature}

The linear SSP example above is an example of a medium with constant
(negative) Gaussian curvature---a hyperbolic space. In general, it is
worthwhile to ask which SSPs yield spaces of constant positive or
negative curvature, because these SSPs serve as models for both
convergent and divergent ray propagation.  Solving the 2d~order
nonlinear differential equation ${cc''-(c')^2=K}$ for constant~$K$
yields, \begin{dbleqnarray} c(z)&=&c_0\cosh (z-z_0)/W;&
  K&=&c_0^2/W^2>0,\nonumber\\ c(z)&=&c_0\sinh (z-z_0)/W;&
  K&=&-c_0^2/W^2<0,\nonumber\\ c(z)&=&c_0\cos (z-z_0)/W;&
  K&=&-c_0^2/W^2<0,\nonumber\\ c(z)&=&c_0\sin (z-z_0)/W;&
  K&=&-c_0^2/W^2<0,\nonumber\\ c(z)&=&c_0+\gamma z;&
  K&=&-\gamma^2\le0,\end{dbleqnarray} of which only the first has
constant positive curvature.

The hyperbolic cosine SSP $c(z) =c_0\cosh({z-z_0})/W\approx
c_0+\texthalf c_0W^{-2}({z-z_0})^2$ is the solution for an acoustic
duct, which traps rays near depth~$z_0$. Regions modeled by hyperbolic
cosine SSPs are described in the literature with increasing
frequency,~\cite{Tolstoy66,Brock77,Tolstoy85,Zhang87,Porter90,Kordich97,Bergman06,Vadov06,Ingard10};
Bergman~\cite{Bergman06} observes that this profile yields constant
curvature. The solution to Jacobi's equation for intrinsic geometric
spreading in Theorem~\ref{thm:jacobi-equation2} equals
${\intrgeomspr(t) =K^{-{1\over2}}\sin K^{1\over2}t}$, with
half-wavelength travel-time between caustics determined by ${\pi
  K^{-{1\over2}}=\pi W/c_0}$
[cf.\ Eq.~(\ref{eq:half-wavelength})]. The range between caustics
equals $c_0$ times the travel-time for~all small initial elevation
angles $\theta_0$; therefore, by invariance, the range between
caustics for~all initial elevation angles necessarily equals~$\pi W$,
yielding the following new and useful
theorem~\cite{Tolstoy66,Brock77,Tolstoy85,Bergman06}:

\begin{theorem}\label{thm:cosh-ssp}Let ${c(z)=c_0\cosh(z-z_0)/W}$ be a
  SSP with hyperbolic cosine profile. Then for any initial elevation
  angle~$\theta_0$, speed~$c_0$, and depth~$z_0$, the range between
  caustics and focusing regions is given by the constant, \begin{equation}
    {\hbox{half-wavelength}\atop\hbox{range}}= \pi c_0K^{-1/2}= \pi
    W. \label{eq:cosh-ssp-half-wavelength}\end{equation}
\end{theorem}

This theorem is illustrated in Fig.~\ref{fig:cosh-ssp} over a large
range of initial elevation angles.  Porter~\cite{Porter90} uses a value
of~${W^{-1}=0.0003}\,\hbox{m}^{-1}$ for his hyperbolic cosine SSP and
observes without explanation that ``the rays refocus perfectly at
distances of about every $10\,$km.'' Indeed,
Theorem~\ref{thm:cosh-ssp} establishes that all rays refocus perfectly
every ${\pi/0.0003=10.47}\,$km, precisely as shown in Porter's Fig.~8
on p.~2020.~\cite{Porter90} This is also verified by the closed-form
functional form of rays with SSP, available from Snell's law,
$\cosh({z-z_0})/W=\sec\theta_0\*\bigl({1-\sin^2\theta_0\cos^2
  (r/W)}\bigr)^{1\over2}$; the property of constant curvature also
yields a simple solutions for the intrinsic geometric spreading
${\intrgeomspr(t)=(c_0/W)\sin(c_0t/W)}$ and both the intrinsic phase
${\dotintrgeomspr/\intrgeomspr =(W/c_0)\cot(c_0t/W)}$ and extrinsic
phase $p/q$ via Eq.~(\ref{eq:c2pq}).  In practice, it is the quadratic
term ${\cosh z\approx1+\texthalf z^2}$ of the hyperbolic cosine
profile that dominates ray behavior in a duct, and ducting regions
with quadratic SSPs and approximately constant positive curvature are
quite common.  This establishes the approximate distance between
caustics in a duct given in Eq.~(\ref{eq:half-wavelength}).

In all other solutions with physical positive speed ${c(z)>0}$, the
Gaussian curvature is a negative constant, and rays diverge from each
other according to the intrinsic geometric spreading ${\intrgeomspr(t)
  =K^{-{1\over2}}\sinh K^{{1\over2}}t}\approx {t+{1\over 6}Kt^3}$,
which equals ${t+{1\over 6}(c_0^2/W^2){\cdot}t^3}$ for the downward
refracting SSP ${c(z)=c_0\sinh(z-z_0)/W}$ and the trigonometric SSPs
${c(z)=c_0\cos(z-z_0)/W}$ with divergence zones. The trigonometric
SSPs of constant negative curvature with convergent ducts (such that
${c''>0}$) are in fact nonphysical because the sound speed must be
negative at such depths; physical ducting behavior with positive sound
speed is described by the first hyperbolic cosine solution. The case
of linear SSPs, whose constant curvature is negative, is treated in
the previous subsection. Finally, SSPs with vanishing constant
curvature ${K\equiv0}$ and intrinsic geometric spreading
${\intrgeomspr(t) =t}$ are obviously given by the SSP with constant
speed~$c_0$.

\subsection{\label{sec:snell}Beam Amplitude and Phase Based Upon Snell's Law}

From Fig.~\ref{fig:snellgeomspread} it is graphically obvious that
the geometric spreading $q(t) =\|(\partial\xb(t;
\theta_0) /\partial\theta_0)_t \|_2$ satisfies the trigonometric
relationships \begin{equation} q =c\intrgeomspr
  =\left(\left({\partial r\over\partial\theta_0}\right)_t^2
    +\left({\partial
        z\over\partial\theta_0}\right)_t^2\right)^{1/2} \\
  =\left({\partial r\over\partial\theta_0}\right)_z\sin\theta
  =-\left({\partial
      z\over\partial\theta_0}\right)_r\cos\theta,
  \label{eq:egstrig}
\end{equation} where $\theta = \atantwo(\dot z,\dot r)$ is the
elevation angle of the ray at point~$t$. This figure is drawn
using an implicit assumption Gauss's lemma, namely that the tangent
vectors $\dot\xb$ and~$(\partial\xb/\partial\theta_0)_t$ are
perpendicular. It is therefore instructive in establishing the
relationship between Snell's law and the intrinsic approach to show
how Gauss's law imply Eqs.~(\ref{eq:egstrig}). An additional benefit
will be an equation for the transverse phase of a Gaussian beam
expressed in Snell's law form.

The first-order Taylor series expansion of the expressions
$r(t;\theta_0)$ and~$z(t;\theta_0)$ yield \begin{equation} \delta r
  =\dot r\,\delta t_{\rm e} + (\partial
  r/\partial\theta_0)_t\,\delta\theta_0, \quad\hbox{and}\quad \delta z
  =\dot z\,\delta t + (\partial
  z/\partial\theta_0)_t\,\delta\theta_0. \end{equation} Fixing the
variables $z$ and~$r$, i.e.\ ${\delta z=0}$ and~${\delta r=0}$
respectively, yields the relationships \begin{equation} (\partial
  r/\partial\theta_0)_z = -\cot\theta\cdot(\partial
  z/\partial\theta_0)_r = (\partial r/\partial\theta_0)_t -
  \cot\theta\cdot(\partial z/\partial\theta_0)_t \label{eq:pdrelns}
\end{equation} between partial derivatives. Gauss's lemma
[Eq.~(\ref{eq:Gausslemma})] provides the additional relationship
\begin{equation} (\partial r/\partial\theta_0)_t +
\tan\theta\cdot(\partial z/\partial\theta_0)_t =0\end{equation}
that along with Eq.~(\ref{eq:pdrelns}) yields
Eq.~(\ref{eq:egstrig}). The conclusion is that the intrinsic
geometric spreading
\begin{equation}\intrgeomspr =c^{-1}(\partial
r/\partial\theta_0)_z\sin\theta =-c^{-1}(\partial
z/\partial\theta_0)_r\cos\theta \label{eq:intrinsicgssnell}
\end{equation} in Snell's law form satisfies Jacobi's equation
[Eq.~(\ref{eq:Jacobi2d})], where $r(z;\theta_0)$ is given by Snell's
law [Eq.~(\ref{eq:snells-law})].

A benefit of this intrinsic viewpoint of Snell's law is a new
expression for the transverse phase of a Gaussian beam, obtained by
direct application of Eq.~(\ref{eq:c2pq}) to
Eqs.\ (\ref{eq:snells-law}) and~(\ref{eq:egstrig}). Differentiating
the first equality of Eq.~(\ref{eq:egstrig}) w.r.t.\ $s$ and applying
the identities ${dz/ds =\sin\theta}$ the Frenet-Serret equations of
section~\ref{sec:ray-euler-lagrange-frenet-serret}
yields \begin{equation} p/q = {1\over c}\left({\partial
    r\over\partial\theta_0}\right)_z^{-1} \left({\partial
    r\over\partial\theta_0}\right)_z'\sin\theta +{c'\over
    c^2}{\cos^2\theta\over
    \sin\theta}, \label{eq:phasesnell}\end{equation} where $({\partial
  r/\partial\theta_0})_z' =(d/dz)({\partial r/\partial\theta_0})_z$ as
usual. These terms are immediately available via
Eq.~(\ref{eq:snells-law}): \begin{eqnarray} \left({\partial
    r\over\partial\theta_0}\right)_z &=& -{\sin\theta_0\over c(z_0)}
  \int_{z_0}^z{c(z_\prime)
    \over\bigl(1-a^2c^2(z_\prime)\bigr)^{3/2}}\,dz_\prime,
  \label{eq:dtsnells-law} \\ \left({\partial
    r\over\partial\theta_0}\right)_z' &=& -{\sin\theta_0\over c(z_0)}
    {c(z) \over\bigl(1-a^2c^2(z)\bigr)^{3/2}}.
  \label{eq:dtdzsnells-law} \end{eqnarray} Likewise, differentiating
the first equality of Eq.~(\ref{eq:intrinsicgssnell}) w.r.t.~$t$
yields ${\dotintrgeomspr/\intrgeomspr =c^2p/q-c'\sin\theta}$, consistent
with Eqs.\ (\ref{eq:ptilde}) and~(\ref{eq:c2pq}).

\subsection{\label{sec:ray-angle}Beam Amplitude and Phase Based
Upon Ray Angle}

The derivation of the phase of a Gaussian beam above leads to another
formulation of the amplitude of a Gaussian beam, which has been used
in Gaussian beam applications.  Eqs.\ (\ref{eq:qde})
and~(\ref{eq:Ndxdsip}) implies that up to second order
in~$\delta\theta_0$, \begin{equation} q(s) =\int_0^s
  (\partial\theta/\partial\theta_0)_{s_\prime}\,ds_\prime,
\end{equation} a formula that has been used by others to compute the
extrinsic geometric spreading. The transverse phase of a Gaussian
beam, up to second-order, is then given by \begin{equation} {cp\over
    q} =\left(\int_0^s \left({\partial\theta\over
    \partial\theta_0}\right)_{s_\prime}\, ds_\prime\right)^{-1}
  \left({\partial\theta\over
    \partial\theta_0}\right)_s \label{eq:phasespoff} \end{equation}
and Eq.~(\ref{eq:extdifftime}).

\begin{figure}[t]
\centerline{\includegraphics[width=.5\columnwidth]{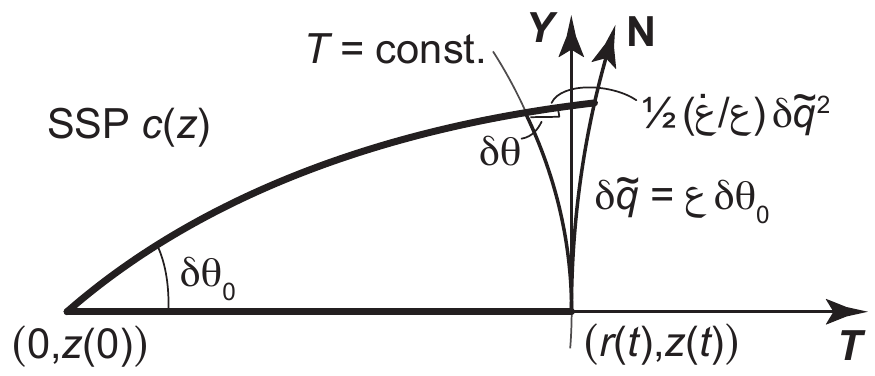}}
\caption{Intrinsic view of ray tube in a horizontally stratified
  medium, in the tangent plane at the point~$\bigl(0,z(0)\bigr)$. The
  azimuthal component, not shown, is perpendicular to the page. All
  distances in the figure represent travel-time, not Euclidean
  distance. Note that the extrinsic ray normal~${\Nb=d\Tb/ds}$---the
  straight line in extrinsic coordinates used to define $\delta q$ in
  Eq.~(\ref{eq:qdef})---is not straight using intrinsic coordinates
  and diverges a small amount from the time-minimizing geodesic in the
  normal direction, shown here as the vector~$\Ybmit$. The extrinsic
  normal vector~$\Nb$ is preferred for numerical computation of the
  differential path time ${\delta t_{\rm e}=\texthalf(p/q)\,\delta\eta^2}$;
  however, the intrinsic normal vector $\Ybmit$, from which the
  differential travel-time is given by $\delta
  t_{\rm i}=\texthalf(c^{-2}\dotintrgeomspr/\intrgeomspr)\,\delta\eta^2
  =\texthalf(p/q-c'\sin\theta),\delta\eta^2$, appears in the intrinsic
  formulation of Gaussian beam. The path time along either
  intrinsic~$\Ybmit$ or extrinsic~$\Nb$ equals ${\delta\tilde
    q=\intrgeomspr\,\delta\theta_0}$ to
  first-order.\label{fig:intrinsic-raytube}}
\end{figure}

\subsection{\label{sec:gb-ampphase}Intrinsic Transverse Amplitude and Phase of a Gaussian beam}

The amplitude and phase of a Gaussian beam has been fully developed in
section~\ref{sec:geometric-spreading} for the computationally
practical case of a Gaussian beam defined extrinsically along a
straight line normal to the ray.  In this section, it is explained how
Jacobi's equation quantifies precisely the same construct
intrinsically, if the extrinsic straight line in ${(r,z)}$-space is
replaced with an intrinsic geodesic normal to the
ray. Fig.~\ref{fig:intrinsic-raytube} illustrates the difference:
because ``straight lines'' in ${(r,z)}$-space are actually curved
intrinsically---they are not time minimizing geodesics---there is a
small difference between the differential path times. This accounts
for the extra term~$c'\sin\theta$ seen in Eqs.\ (\ref{eq:ptilde}),
(\ref{eq:c2pq}), and~(\ref{eq:intrayampphase}). Though computationally
impractical, the following new theorem connects Jacobi's equation with
the differential path time along rays normal to the ray at the center
of the Gaussian beam.

\begin{theorem} {\rm(Intrinsic transverse phase of a Gaussian beam, $2$\hbox{-}d
case)}\quad \label{thm:gbphase2} Let $\delta t_{\rm i}$ be the travel-time
  difference between the surface of constant travel-time $t$ and an
  intrinsic distance $\delta\mu$ along the intrinsic transverse plane
  (i.e.\ defined by geodesics) to the ray at point $\xb(t;\theta_0)$,
  and let $\intrgeomspr(t)$ be the solution to Jacobi's equation
  [Eq.~(\ref{eq:Jacobi2d})] along the ray. Then \begin{equation}
    \delta t_{\rm i} ={1\over2}{\dotintrgeomspr\over\intrgeomspr}\,\delta\mu^2
    +O(\delta\mu^3).\label{eq:dellambdaquad}\end{equation}
\end{theorem}

The proof of Theorem~\ref{thm:gbphase2} involves another application
of Gauss's lemma and a Taylor series expansion about
$\xb(t;\theta_0)$, and is comparable to the proof of
Eq.~(\ref{eq:extdifftime}) above.  Consider the change of elevation
angle $\theta(t) =\atantwo(\dot z,\dot r)$ of the ray: let
$\delta\theta$ be the difference between $\theta$ and the slope of the
ray $\xb(t; {\theta_0+\delta\theta})$ intersecting the transverse
plane at distance $\delta\mu =\intrgeomspr\,\delta\theta_0$ along a
geodesic emanating perpendicularly from the ray with initial unit
tangent vector~$\Ybmit$, i.e.\ ${g(\Ybmit,\Ybmit)=1}$
and~${g(\Tbmit,\Ybmit)=0}$.  This definition of~$\delta\theta$ equates
to the relation \begin{equation} \delta\theta =\langle\Ybmit,
  \tau_{\delta\mu}^{-1} \dot\xb(t;\theta_0 +\delta\theta)
  -\dot\xb(t;\theta_0)\rangle,
  \label{eq:deltathetadef} \end{equation} where
$\tau_{\delta\mu}^{-1}$ is the inverse parallel translation along the
geodesic between the points $\xb(t;\theta_0)$
and~$\xb(t;{\theta_0+\delta\theta})$ in direction $\Ybmit$. A
straightforward Taylor series analysis of the differential equations
describing parallelism~\cite{Helgason78} shows that for any
arbitrary vector field $\Xbmit(\delta\mu)$,
$\tau_{\delta\mu}^{-1}\Xbmit(\delta\mu) =\Xbmit(0)
+O(\delta\mu^2)$---the proof involves so-called ``normal coordinates''
defined so that at the single point where $\delta\mu=0$, $g_{ij}(0)
=\delta_{ij}$ and rays in the direction $\Ybmit$ have the coordinates
$(t Y^1, t Y^2)$, implying that $\Gamma^k_{ij}(0)=0$.  The
implication of this local analysis about the point
$\xb(t;\theta_0)$ is that
\begin{equation} \delta\theta =\dotintrgeomspr\,\delta\theta_0
+O(\delta\theta_0^2) \label{eq:deltatheta1} \end{equation} because, up
to second order, \begin{equation} \dot\xb(t;{\theta_0+\delta\theta})
  = (d/dt)\bigl(\xb(t;{\theta_0}) +\intrgeomspr(t)\Ybmit(t)\bigr)
  = \dot\xb(t;\theta_0) +\dotintrgeomspr\Ybmit
  +\intrgeomspr\nabla_\Tbmit\Ybmit = \dot\xb(t;\theta_0)
  +\dotintrgeomspr\Ybmit \end{equation} by the parallelism of~$\Ybmit$
along the ray.  Now the change in slope $\delta t_{\rm i}$ of the ray is equal
to the slope of the quadratic function $\delta t_{\rm i} ={(1/2)Q_{\rm
    i}\,\delta\mu^2+O(\delta\mu^3)}$ for some coefficient~$Q_{\rm i}$
that is implied by Gauss's lemma. The slope of this function
at~$\delta\mu$ equals
\begin{equation} \delta\theta +O(\delta\theta^3) =\tan\delta\theta
  =d(\delta t_{\rm i})/d(\delta\mu) =Q_{\rm i}\,\delta\mu =Q_{\rm
    i}\intrgeomspr\,\delta\theta_0. \label{eq:ddellambdaddelmu}\end{equation}
Equating Eqs.\ (\ref{eq:deltatheta1}) and~(\ref{eq:ddellambdaddelmu}),
we conclude that ${Q_{\rm i} =\dotintrgeomspr/\intrgeomspr}$,
establishing the theorem.

The quadratic coefficient $\dotintrgeomspr/\intrgeomspr$ for the phase
is physically consistent with two easily imagined examples.  For flat
space where the Gaussian curvature vanishes, $\intrgeomspr(t)\equiv
t$, $\dotintrgeomspr(t)\equiv1$, and the wavefront of constant~$t$ is
a circle of radius~$t$, in which case the gap size equals $\delta t_{\rm i}
-{t\sec\delta\theta_0-t} ={(t/2)(\delta\theta_0)^2+O(h^4)}
={(1/2)(1/t)(\delta\mu)^2+O(h^4)}$, as predicted by
Theorem~\ref{thm:gbphase2}.  For the second example, consider acoustic
propagation in a space of constant positive curvature $K=R^{-2}$,
e.g.\ S-waves emanating from a (geologically unlikely) earthquake on
the north pole of the earth.  In this case the rays are all great
circles on the sphere that intersect at the north pole, i.e.\
longitudinal lines. At every point around the equator, equivalently,
in the plane transverse to every ray at the equator, all rays have
equal travel-time to the north pole; therefore, $\delta t_{\rm i} =0$ around
the equator.  Indeed, solving Jacobi's equation
[Eq.~(\ref{eq:Jacobi2d})] yields $\intrgeomspr(t) =\sin(t/R)$ for
which $\dotintrgeomspr ={\cos({1\over2}\pi) =0}$ at the equator, an
easily obtained intuitive result that is again consistent with
Theorem~\ref{thm:gbphase2}.

\section{\label{sec:conc}Conclusions}

Paraxial ray theory is an essential and useful component of modern
acoustic modeling. There are deep and important connections between
the classically derived paraxial ray equations for the amplitude and
phase along a Gaussian beam, and the second-order variation---called
Jacobi's equation---along geodesics encountered in differential
geometry. This paper demonstrates how known results in paraxial ray
theory correspond to their counterparts in differential geometry, and
shows how both new equations and new insights into the properties of
acoustic rays are obtained from an intrinsic, differential geometric
point of view. It is shown how the intrinsic Gaussian curvature affect
the spreading of Gaussian beams, and how the specific form of this
intrinsic curvature [${K=cc''-(c')^2}$ for a horizontally stratified
  medium with SSP~$c(z)$] allows one to easily compute the distance
between caustics within a duct, as well as the geometric spreading for
either a duct or a region with linear SSP. These results allow the
introduction of SSPs yielding constant Gaussian curvature, which serve
as model spaces for both convergent acoustic ducts (positive
curvature), divergence zones (negative curvature), and non-refractive
spreading (zero curvature).  It is proved for the model of a
hyperbolic cosine SSP, the range between caustics is constant for~all
rays.  Intrinsic versions of the amplitude and Gaussian beams are
introduced. The intrinsic geometric spreading is shown to be
equivalent to its extrinsic counterpart after scaling by the
position-dependent sound speed. The intrinsic phase of a Gaussian
beam, which is the phase along geodesics, not straight lines,
emanating in a normal direction from the ray, is shown to be
equivalent up to a small additive term to the phase of a paraxial ray
as it is typically defined.  Because the differential geometric
approach is quite general, all results may be generalized to three
dimensions, and in all cases, the connection is made with known
results derived from either Snell's law, Hamilton's equations, or the
Frenet-Serret formulae. The results may also be applied to other
special cases of applied interest, such as a spherically stratified
sound speed.

\section*{Acknowledgment}
The author thanks Arthur Baggeroer for his encouragement and critiques
of this paper.

\let\bibliographysize=\footnotesize 

\bibliography{strings,refs}

\endgroup 

\end{document}